\documentclass[sigconf]{acmart}

\usepackage{algorithm}
\usepackage{algpseudocode}
\usepackage{makecell}
\usepackage{amsmath}
\usepackage{booktabs}
\usepackage{multirow}
\usepackage[table]{xcolor}   % 注意要带 [table]
\newcommand{\tblhead}{\rowcolor{black!5}}   % 表头浅灰底

\usepackage[framemethod=TikZ]{mdframed}
\usepackage{etoolbox}  % 用来 patch 环境

\BeforeBeginEnvironment{algorithmic}{%
  \begin{mdframed}[
    backgroundcolor=black!2,  % 很浅很浅的灰
    linecolor=black!15,       % 细浅边框
    roundcorner=3pt,
    skipabove=\smallskipamount,
    skipbelow=\smallskipamount,
    innerleftmargin=6pt,
    innerrightmargin=6pt,
    innertopmargin=6pt,
    innerbottommargin=6pt
  ]%
}
\AfterEndEnvironment{algorithmic}{%
  \end{mdframed}
}

\AtBeginDocument{%
  }

\copyrightyear{2026}
\acmYear{2026}
\setcopyright{cc}
\setcctype{by}
\acmConference[WWW '26]{Proceedings of the ACM Web Conference 2026}{April 13--17, 2026}{Dubai, United Arab Emirates}
\acmBooktitle{Proceedings of the ACM Web Conference 2026 (WWW '26), April 13--17, 2026, Dubai, United Arab Emirates}
\acmPrice{}
\acmDOI{10.1145/3774904.3792996}
\acmISBN{979-8-4007-2307-0/2026/04}

\acmSubmissionID{353}
\begin{document}

\title{Egocentric Co-Pilot: Web-Native Smart-Glasses Agents for Assistive Egocentric AI}

\author{Sicheng Yang}
\orcid{0009-0009-2018-0604}
\affiliation{%
  \department{\mbox{Shenzhen International Graduate School}}
  \institution{Tsinghua University}
  \city{Shenzhen}
  \country{China}}
\email{yangsc25@mails.tsinghua.edu.cn}

\author{Yukai Huang}
\orcid{0009-0009-5725-5884}
\affiliation{%
  \institution{Independent Researcher}
  \city{London}
  \country{United Kingdom}}
% \email{k.study168@gmail.com}
\email{u06530032@alum.ccu.edu.tw}

\author{Weitong	Cai}
\orcid{0000-0001-7726-4387}
\author{Shitong	Sun}
\orcid{0000-0003-1825-655X}
\affiliation{%
  \institution{Queen Mary University of London}
  \city{London}
  \country{United Kingdom}}
\email{weitong.cai@qmul.ac.uk}
\email{shitong.sun@qmul.ac.uk}

\author{Fengyi Fang}
\orcid{0000-0002-1082-8368}
\author{You	He}
\orcid{0000-0002-2942-1699}
\authornote{Corresponding author.}
\affiliation{%
  \department{\mbox{Shenzhen International Graduate School}}
  \institution{Tsinghua University}
  \city{Shenzhen}
  \country{China}}
\email{fangfy22@tsinghua.org.cn}
\email{heyou@mail.tsinghua.edu.cn}

\author{Yiqiao Xie}
\orcid{0009-0007-8566-1607}
\author{Jiankang Deng}
\orcid{0000-0002-3709-6216}
\affiliation{%
  \institution{Imperial College London}
  \city{London}
  \country{United Kingdom}}
\email{yx2722@ic.ac.uk}
\email{j.deng16@imperial.ac.uk}

\author{Hang Zhang}
\orcid{0000-0003-0115-387X}
\affiliation{%
  \institution{Independent Researcher}
  \city{London}
  \country{United Kingdom}}
  \email{hz459@cornell.edu.cn}

\author{Jifei Song}
\orcid{0000-0002-3381-6685}
\affiliation{%
  \institution{University Of Surrey}
  \city{Guildford}
  \country{United Kingdom}}
\email{j.song@qmul.ac.uk}

\author{Zhensong Zhang}
\orcid{0009-0001-7911-7564}
\affiliation{%
  \institution{Independent Researcher}
  \city{London}
  \country{United Kingdom}}
\email{zhensongzhang@hotmail.com}

\renewcommand{\shortauthors}{Sicheng Yang et al.}

\begin{abstract}
What if accessing the web did not require a screen, a stable desk, or even free hands? For people navigating crowded cities, living with low vision, or experiencing cognitive overload, smart glasses coupled with AI agents could turn the web into an always-on assistive layer over daily life. We present \emph{Egocentric Co-Pilot}, a web-native neuro-symbolic framework that runs on smart glasses and uses a Large Language Model (LLM) to orchestrate a toolbox of perception, reasoning, and web tools. An egocentric reasoning core combines Temporal Chain-of-Thought with Hierarchical Context Compression to support long-horizon question answering and decision support over continuous first-person video, far beyond a single model's context window. On top of this, a lightweight multimodal intent layer turns noisy speech and gaze into structured, tool-ready commands without relying on a single monolithic model. We further implement and evaluate a cloud-native WebRTC pipeline based on LiveKit, integrating streaming speech, video, and control messages into a single web-standard channel that serves both smart-glasses clients and browser-based playgrounds. In parallel, we deploy an on-premise WebSocket baseline, exposing concrete trade-offs between local inference and cloud offloading in terms of latency, mobility, and resource use. Experiments on Egolife and HD-EPIC demonstrate competitive or state-of-the-art egocentric QA performance, and a human-in-the-loop study on smart glasses shows higher task completion and user satisfaction than leading commercial baselines. Taken together, these results indicate that web-connected egocentric co-pilots can be a practical path toward more accessible, context-aware assistance in everyday life. By grounding operation in web-native communication primitives and modular, auditable tool use, Egocentric Co-Pilot offers a concrete blueprint for assistive, always-on web agents that support education, accessibility, and social inclusion for people who may benefit most from contextual, egocentric AI. 
% Upon acceptance, we will release our code, model checkpoints, and detailed fine-tuning and evaluation scripts.
Our code, fine-tuned models are available at \href{https://github.com/YoungSeng/Egocentric-Co-Pilot}{here}.
\end{abstract}




\begin{teaserfigure}
\centering
  \includegraphics[width=0.67955\linewidth]{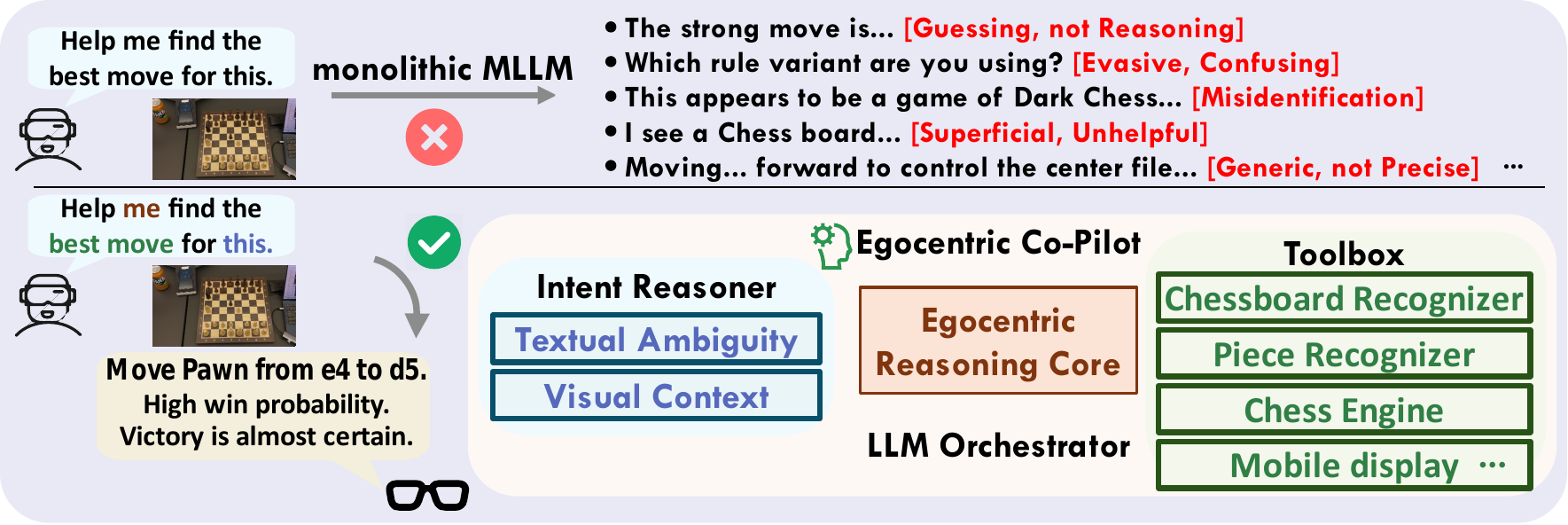}
  % \caption{Overview of the Egocentric Co-Pilot framework, contrasted with a monolithic approach. Top: A monolithic MLLM struggles with a specialized reasoning task (e.g., a strategy board game), providing inaccurate and evasive answers. Bottom: Our framework employs an LLM orchestrator that leverages a toolbox of specialized neuro-symbolic modules. It successfully interprets the user's request and invokes the perception modules to generate board state, which is then solved by a dedicated game engine, leading to a precise and actionable solution.}
  \caption{Overview of Egocentric Co-Pilot vs. Monolithic MLLM. Top: Monolithic MLLMs struggle with specialized reasoning task (e.g., a strategy board game), often providing evasive answers. Bottom: Our framework uses an LLM orchestrator that leverages a toolbox of specialized neuro-symbolic modules. It successfully interprets the user's request and invokes the perception modules to generate board state, which is then solved by a dedicated game engine, leading to a precise and actionable solution.}        % inaccurate and
  \Description{Egocentric Co-Pilot pipeline contrasted with a monolithic model.}
  \label{fig:teaser}
\end{teaserfigure}

\maketitle

\section{Introduction}
Combining powerful, compact hardware and large language models (LLMs) enables a new generation of AI-powered personal assistants~\cite{DBLP:journals/corr/abs-2308-13561}. While autonomous web agents have changed how AI systems interact with digital interfaces~\cite{DBLP:conf/nips/DengGZCSWSS23, DBLP:conf/iclr/ZhouX0ZLSCOBF0N24}, extending these capabilities into the physical world remains challenging. Among potential platforms, smart glasses are particularly suitable, offering a hands-free, always-on interface for overlaying digital information onto the physical world~\cite{DBLP:journals/corr/abs-2407-20242, DBLP:journals/corr/abs-2506-22355, DBLP:journals/imwut/HuangXPYZHCCWNLLFLYQWW25, DBLP:conf/icchp/MuchaCEKT24, anonymous2025wearvox, zhang2024empowering}, and can in principle make web content available to people who struggle with small screens, complex menus, or visually demanding layouts. In practice, many of these assistants are accessed through web-based services and applications, making the web the primary distribution channel and governance surface for their impact. The objective is not merely information retrieval, but the creation of a \emph{co-pilot} for human cognition: an agent that operates from the user's first-person (egocentric) viewpoint to understand intent and provide proactive assistance in everyday tasks such as reading nutrition labels, following multi-step instructions, or tracking appointments \cite{DBLP:journals/corr/abs-2501-01428, DBLP:journals/tce/DiWSFZML25, tomavsev2020ai}.

Beyond technical performance, assistants meant for broad, beneficial use must integrate with web-native communication and security primitives (standard web protocols, cloud APIs, and browser-based clients) so that they can be deployed responsibly and efficiently at scale~\cite{DBLP:conf/nips/PatilZ0G24, DBLP:journals/tors/DiSMGLW26}. For the kinds of users we target—including people with low vision, attention or memory challenges, or limited mobility—such agents should prioritize reliability, privacy, and reduced cognitive load over engagement or novelty. In practical deployments, this translates into predictable behavior, clear signaling when the system is uncertain, and explicit controls over what information is streamed or stored. However, realizing this vision requires overcoming several fundamental challenges.
First, real-world interactions are inherently ambiguous. For example, a simple deictic command like ``analyze this'' must be grounded in a cluttered visual scene to resolve its reference, a task that demands robust multimodal reasoning \cite{DBLP:journals/corr/abs-2411-14466}. Second, no single AI model can solve all problems effectively \cite{DBLP:journals/corr/abs-2506-03569}. Many tasks demand a combination of robust perception, where neural networks excel, and precise symbolic reasoning or tool use, such as planning moves in a game or calling a calendar API. Current end-to-end models often lack the precision needed for these specialized tasks \cite{DBLP:journals/nn/YuYLWP23, weng2023agent}. Finally, the continuous nature of egocentric data streams poses a problem for models with finite context windows, which struggle to maintain long-range dependencies and temporal context \cite{DBLP:conf/iclr/YeZD0LL0Y0GLY25}. 

In this paper, we argue that an effective egocentric assistant should be built on a modular, neuro-symbolic architecture orchestrated by a central LLM. We introduce the \textbf{Egocentric Co-Pilot}, a framework designed to connect human intent with a set of specialized tools and web-accessible services. Instead of relying on a single model, our framework uses an LLM as a reasoning engine to interpret the user's multimodal commands. It first clarifies intent through interactive dialogue and visual grounding. Then, it generates execution plans by selecting and invoking the most suitable tools like neural perception modules, symbolic reasoners, or external web APIs. This hybrid approach combines the contextual understanding of LLMs with the precision of specialized modules, while exposing a web-native interface that can run on resource-constrained devices and browser-based clients in everyday settings.        % ---to complete the task

Our core contribution is a framework that addresses the above challenges in a way that is compatible with socially beneficial web deployments. 
Specifically, we make the following contributions:
\begin{enumerate}
    \item \textbf{A framework for combining specialized tools.} We propose a neuro-symbolic framework that uses a Large Language Model (LLM) to coordinate on-device modules and web-based services through a lightweight, web-friendly protocol (MCP). This makes it possible to expose a rich ecosystem of perception, reasoning, and assistive web APIs to resource-constrained devices such as smart glasses.

    \item \textbf{A module for understanding ambiguous commands.} To handle unclear user requests, we design a module that clarifies intent before acting. For vague text, it asks follow-up questions. For ambiguous visual input, it uses a 3D ray-casting method to determine what a user is pointing at, ensuring commands are interpreted correctly. This conservative, user-centric behavior is especially important in safety-critical or socially impactful contexts, where avoiding harmful misunderstandings matters more than raw throughput.

    \item \textbf{A module for reasoning over long videos.} To process continuous egocentric video, we develop a method for context management that combines \textit{Temporal Chain-of-Thought (T-CoT)} for detailed, short-term reasoning with \textit{Hierarchical Context Compression (HCC)} for long-term memory. This allows the system to use information from periods longer than the model's standard context window, while operating on streams captured by resource-constrained devices.      % web-connected, 

    % (3) A module for reasoning over long videos. To process continuous egocentric video, we develop a method for context management that combines Temporal Chain-of-Thought (T-CoT) for detailed, short-term reasoning with Hierarchical Context Compression (HCC) for long-term memory. This allows the system to use information from periods longer than the model’s standard context window, while operating on streams captured by web-connected, resource-constrained devices, and it exposes intermediate summaries that can be inspected or adapted for different applications.

    \item \textbf{A complete system with real-world evaluation.} We build our framework into a working system on smart glasses with a web-native backend and test it extensively, achieving strong results on egocentric QA benchmarks (Egolife, HD-EPIC). Furthermore, we demonstrate its practical value in a user study, where it significantly outperforms leading commercial systems on real-world tasks that emphasize constructive, everyday assistance rather than engagement-only use cases.      % standard

    % (4) A complete system with real-world evaluation. We build our framework into a working system on smart glasses with a web-native backend and test it extensively. It achieves strong results on standard egocentric QA benchmarks (Egolife, HD-EPIC), and our ablations quantify the contribution of each component in the reasoning core. More importantly, we demonstrate its practical value in a human-in-the-loop study, where it significantly outperforms leading commercial systems on real-world tasks across three representative categories—simple read-and-point assistance, strategy-board tutoring, and open-ended everyday support—chosen to reflect recurring, socially meaningful use cases rather than entertainment-only interactions.

\end{enumerate}

\section{Related Work}

\paragraph{Egocentric Artificial Intelligence.}
Egocentric AI has evolved from foundational tasks like action recognition and hand--object interaction \cite{DBLP:journals/corr/abs-1804-02748, DBLP:conf/cvpr/PerrettDSEPPLGB25, DBLP:conf/wacv/ShiotaTKSA24, DBLP:journals/corr/abs-2312-15719}, and visual sentiment analysis~\cite{DBLP:journals/tmm/RuanZWXLC24} to complex reasoning with Multimodal Large Language Models (MLLMs) \cite{DBLP:journals/corr/abs-2504-04550, DBLP:journals/corr/abs-2503-09143, DBLP:journals/corr/abs-2509-07447}. MLLMs enable high-level, open-ended tasks such as dense captioning and question answering, evaluated on benchmarks like Ego4D and Egolife \cite{DBLP:journals/corr/abs-2504-04550, DBLP:conf/cvpr/GraumanWBCFGH0L22, yang2025egolife, DBLP:conf/acl/TianZ0025, DBLP:conf/cvpr/DiX24}. However, processing long-form video remains a significant challenge, limited by the computational cost and context windows of transformer-based models \cite{DBLP:conf/iclr/YeZD0LL0Y0GLY25, DBLP:journals/corr/abs-2312-05269}. Common solutions involve converting video to textual logs, hierarchical modeling, or summarization \cite{DBLP:conf/iclr/YeZD0LL0Y0GLY25, DBLP:journals/corr/abs-2501-00574, DBLP:journals/corr/abs-2409-06299}. 
Most of this work focuses on offline analysis, including recent hierarchical-retrieval agents that repeatedly traverse long egocentric logs. In contrast, our system targets real-time, continuous lifelogging on smart glasses, which directly motivates our Hierarchical Context Compression (HCC) method for efficient context management.
% While prior work has focused on offline analysis, including recent hierarchical-retrieval agents that repeatedly traverse long egocentric logs, our system is built for real-time, continuous lifelogging on smart glasses. This application's need for efficient, streaming-friendly context management motivates our proposed dual-horizon approach that combines Temporal Chain-of-Thought (T-CoT) with Hierarchical Context Compression (HCC).

\paragraph{LLM-driven Agents and Tool Use.}
LLMs are now widely used as the central controller for autonomous agents \cite{DBLP:journals/fcsc/WangMFZYZCTCLZWW24, DBLP:journals/corr/abs-2504-04471, he2025intelligent}, enabling them to reason \cite{DBLP:journals/corr/abs-2506-13654}, plan, and interact with external tools through a thought--action loop \cite{DBLP:conf/nips/SchickDDRLHZCS23, DBLP:conf/icml/ZhengGK0024}. While frameworks like LangChain and AutoGPT have simplified development, significant challenges remain in reliability, long-horizon planning, and generalization \cite{topsakal2023creating, DBLP:conf/iclr/YaoZYDSN023}. Much of this research has focused on agents in digital domains, such as web navigation \cite{DBLP:journals/corr/abs-2503-16416, DBLP:conf/acl/HeYM0D0L024, DBLP:conf/iclr/ZhouX0ZLSCOBF0N24, DBLP:conf/nips/DengGZCSWSS23, DBLP:conf/emnlp/YoranAMBPB24}. More recently, researchers have started applying LLMs to physically embodied agents for high-level robotic task planning, where language commands are grounded in the physical world \cite{DBLP:conf/icra/AroraSSDBBJSK24, DBLP:journals/corr/abs-2204-01691, DBLP:conf/icra/ChenA0ZRF24, DBLP:conf/icra/SinghBMGXTFTG23, DBLP:journals/corr/abs-2504-00907}. Our work addresses a specific area within this embodied agent research: non-robotic, wearable agents designed to augment, rather than replace, human action. This ``Egocentric Co-Pilot'' acts as a collaborative partner with the user \cite{DBLP:journals/corr/abs-2407-20242, DBLP:journals/corr/abs-2506-22355}. To facilitate this human--agent collaboration, we introduce the Model-Context Protocol (MCP), a lightweight protocol designed for the real-time, edge--cloud coordination necessary in such a human-augmenting system.

\paragraph{Neuro-Symbolic Systems.}
Neuro-symbolic systems combine neural perception with symbolic reasoning, harnessing the advantages of both approaches. This synthesis helps mitigate their respective weaknesses: the opaque, black-box nature of neural networks and the fragility of symbolic methods when dealing with noisy data \cite{DBLP:journals/nn/YuYLWP23, 10.1093/nsr/nwac035, barnes2024natural}. Much of the recent work in this area has focused on systems that learn from unstructured data while being constrained by explicit symbolic knowledge \cite{DBLP:journals/corr/abs-2501-05435, DBLP:journals/nca/BhuyanRTS24, DBLP:journals/corr/abs-2409-19250}. 
% Our work follows this direction; we employ a neural module to ground a symbolic reasoner by translating raw perceptual input into a structured state representation. However, our approach is distinguished by its architecture. Instead of building a monolithic system with tightly coupled components, we design our neuro-symbolic pipeline as a discrete, callable tool \cite{DBLP:journals/corr/abs-2401-01040, weng2023agent}. 
Our work follows this direction: we employ a neural module to ground a symbolic reasoner by translating raw perceptual input into a structured state representation. Rather than building a monolithic system with tightly coupled components, we package the neuro-symbolic pipeline as a discrete, callable tool \cite{DBLP:journals/corr/abs-2401-01040, weng2023agent}.
This tool is then orchestrated by a Large Language Model (LLM), which results in a hierarchical and modular neuro-symbolic architecture \cite{DBLP:journals/corr/abs-2407-08516, DBLP:journals/corr/abs-2502-20843, DBLP:journals/corr/abs-2503-07148}. This design is consistent with the recent trend of building LLM-based agents that compose specialized modules to solve complex tasks \cite{DBLP:conf/nips/Yi0G0KT18}.

\paragraph{Multimodal Intent Disambiguation.}
Research in intent clarification has progressed from structured dialogue \cite{DBLP:conf/chi/ZhangAHB25, DBLP:journals/corr/abs-2503-01940, li5168033mdsd} to LLM-driven methods \cite{DBLP:conf/acl/ZhangQDHLLJLC24, DBLP:conf/acl/QianHZDQCZZL0024}, yet remains largely confined to disembodied, text-only scenarios \cite{DBLP:conf/wsdm/DammuAP25, DBLP:conf/emnlp/Zhang0RNC24, DBLP:journals/corr/abs-2505-11533}. 
This limitation is particularly acute in egocentric vision (EGV) \cite{DBLP:journals/corr/abs-2503-15275}, where user input is inherently noisy and ambiguous \cite{DBLP:conf/iccvw/Fan19, seth2025egoillusion}, especially for critical modalities like pointing gestures which are often passively processed or oversimplified \cite{DBLP:conf/icra/Das21, DBLP:conf/cvpr/HuangLZJ16, DBLP:conf/cvpr/ManeWSSSM25}. 
This challenge is compounded by the unreliability of even state-of-the-art VLMs \cite{DBLP:journals/corr/abs-2410-21276, google2025gemini25pro, xai2025grok3} for the precise spatial reasoning required, often leading to hallucinations \cite{DBLP:conf/emnlp/MouselinosMM24, DBLP:conf/siggrapha/Feng0WLW24, DBLP:conf/acl/HuangQQLJXW25}. 
% While hybrid architectures that combine LLM reasoning with specialized modules \cite{DBLP:conf/emnlp/0001BZGVJ024, DBLP:conf/naacl/SharmaDKZP25, DBLP:journals/corr/abs-2502-03671} are a promising direction, a key research frontier lies in designing proactive, iterative frameworks that actively guide users to resolve multimodal ambiguity in real time, shifting from the passive analysis of flawed data towards more robust and intuitive interactive systems \cite{DBLP:conf/iclr/LuYQCLWWCZLLWLL25, DBLP:journals/corr/abs-2506-05904}.
Hybrid architectures that combine LLM reasoning with specialized modules \cite{DBLP:conf/emnlp/0001BZGVJ024, DBLP:conf/naacl/SharmaDKZP25, DBLP:journals/corr/abs-2502-03671} are a promising direction. In our work, we take a step toward proactive, iterative frameworks that guide users to resolve multimodal ambiguity in real time, rather than passively analyzing noisy input after the fact \cite{DBLP:conf/iclr/LuYQCLWWCZLLWLL25, DBLP:journals/corr/abs-2506-05904}.
\section{Methodology}

\subsection{Egocentric Reasoning Core}
\label{sec:brain}

\begin{figure}[!t]
    \centering
    \includegraphics[width=0.95\linewidth]{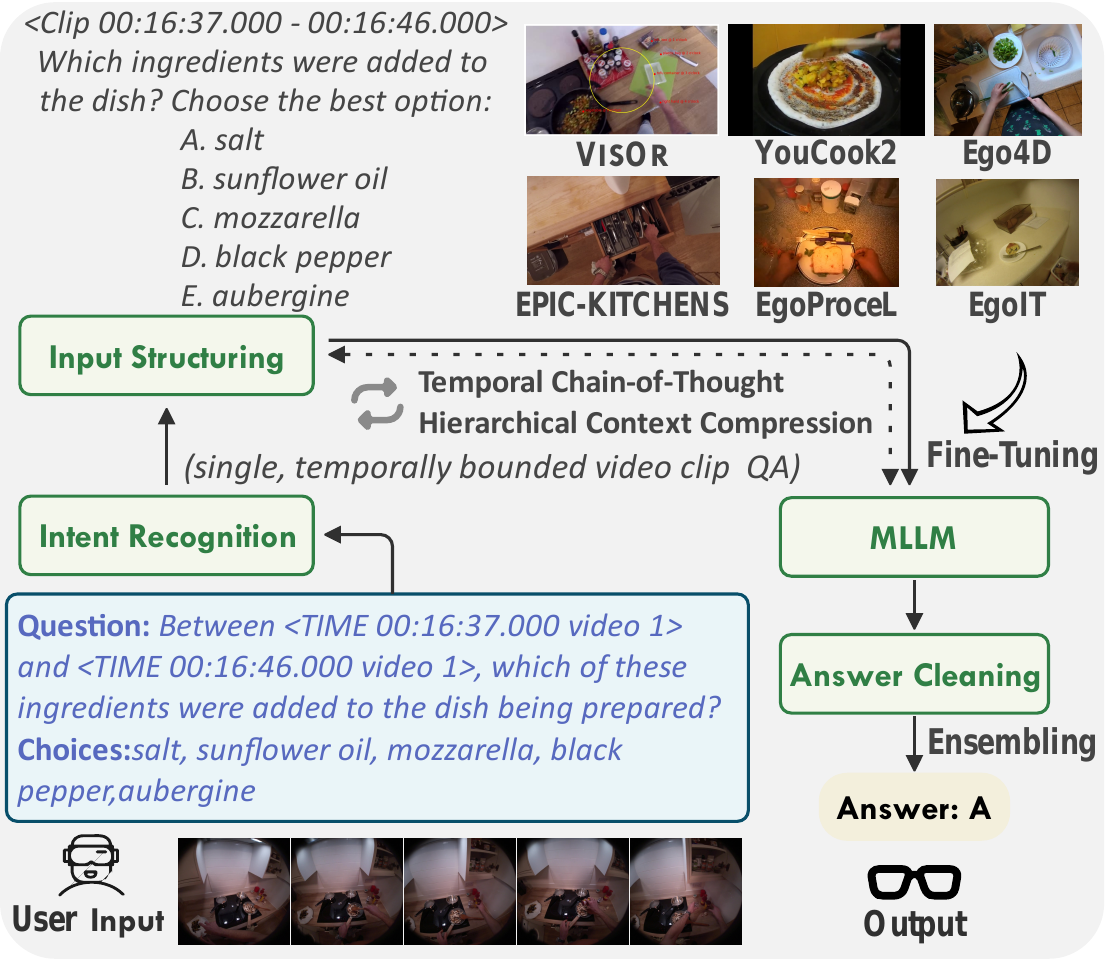}
    \caption{Our Reasoning Core pipeline. It integrates Temporal Chain-of-Thought (T-CoT) for short-term analysis and Hierarchical Context Compression (HCC) for long-term memory. The figure illustrates the T-CoT path where our fine-tuned MLLM processes a temporally bounded query.}
    % \caption{The operational pipeline of our Reasoning Core. It orchestrates context construction using Temporal Chain-of-Thought (T-CoT) for detailed, short-term analysis and Hierarchical Context Compression (HCC) for long-term memory. 
    % The figure illustrates the T-CoT path for a temporally bounded query, which is processed by our fine-tuned MLLM to produce the final output.}
    \label{fig:brain-pipeline}
\end{figure}

At the heart of our framework is the Reasoning Core, an MLLM-based engine designed to process continuous egocentric video streams and answer user queries. The engine's foundation is a unified, chronologically sorted event log $\mathcal{E}$ that integrates dense egocentric video narrations and spoken user queries transcribed via Automatic Speech Recognition (ASR). Each event $e_i = (t_i, m_i, c_i)$ records a timestamp, modality (visual or spoken), and normalized content, giving the system a compact but semantically rich representation of first-person experience.

Upon receiving a user query $Q$, the Reasoning Core initiates a dynamic, multi-stage pipeline to construct the optimal context for the MLLM (Figure~\ref{fig:brain-pipeline}). A key innovation of our approach lies in its dual-level strategy for handling both short-term and long-term temporal dependencies. The process begins by analyzing the query's intent to determine its temporal scope. (1) For fine-grained analysis of recent events or specific time-bounded segments, we employ Temporal Chain-of-Thought (T-CoT) tactics \cite{DBLP:conf/nips/Wei0SBIXCLZ22, yang2026optimizingmultimodalllmsegocentric}. T-CoT programmatically selects a narrow temporal window around relevant timestamps, narrates or clips the corresponding segments, and orders them into a coherent local storyline. This isolates the most pertinent information needed to address the immediate query. (2) For long-term reasoning that spans beyond the MLLM's native context window, we activate Hierarchical Context Compression (HCC). The historical log is partitioned into temporal chunks, each summarized by a smaller text-only model into a short, query-aware description. HCC then chooses the most relevant summaries and prepends them to the T-CoT context, providing long-range awareness without exceeding the model context budget.

% The final reasoning is performed by a Multimodal Large Language Model (MLLM), $\mathcal{M}_{vqa}$ \cite{DBLP:journals/corr/abs-2502-13923}, which we fine-tune on egocentric kitchen and daily-activity datasets (e.g., EPIC-KITCHENS and related corpora) to improve first-person understanding. We standardize multiple-choice options, rewrite under-specified questions into explicit, viewpoint-grounded prompts, and apply a regex-based parser to extract the final choice from free-form model outputs. For evaluation, we ensemble predictions over several prompt variants via majority voting. Together, these steps turn long, noisy egocentric streams into a structured context that is both tractable for current MLLMs and expressive enough for long-horizon reasoning.

The final reasoning is performed by a Multimodal Large Language Model
(MLLM), $M_{\text{vqa}}$~\cite{DBLP:journals/corr/abs-2502-13923}, which we fine-tune on egocentric
kitchen and daily-activity datasets (e.g., EPIC-KITCHENS). We standardize
multiple-choice options, rewrite under-specified questions into explicit
viewpoint-grounded prompts, and use a regex-based parser to extract
the final choice from free-form outputs. For robustness, we ensemble
predictions from several prompt variants via majority voting. Together,
these steps turn long egocentric streams into a tractable context for
long-horizon reasoning. 

\subsection{LLM-Orchestrated Neuro-Symbolic Execution}
\label{sec:our_approach_tools}

To translate the high-level understanding of the Reasoning Core into concrete actions, we use a modular, tool-based model. Unlike monolithic MLLMs handling perception, symbolic computation, and web interaction within a single forward pass, our architecture treats capabilities as callable tools that can be composed on demand.

This orchestration is implemented via the MCP, a lightweight interface that exposes tools as JSON-schema-described functions. 
The execution follows a standard retrieval-action loop: the LLM discovers available tools, formulates a plan, executes tool calls via MCP, and synthesizes the final response. The formal pseudocode for this generalized loop is provided in Appendix~\ref{apx:gen_loop}.
Crucially, MCP is designed to run over standard web channels (e.g., WebRTC data channels or HTTPS backends), so the same tool ecosystem can serve both wearable clients and browser-based users, and can be audited or sandboxed using existing web governance mechanisms.

% \begin{algorithm}[t]
% \small
% \caption{Generalized LLM-Orchestrated Execution Loop}
% \label{alg:generalized_loop}
% \begin{algorithmic}[1]
% \Require User Query $Q$, Multimodal Context $C_{mm}$
% \State $T_{available} \gets \text{MCP.ListTools}()$ \Comment{Discover available tools}
% \State $tool_{plan} \gets \text{LLM.GeneratePlan}(Q, C_{mm}, T_{available})$ \Comment{LLM formulates a tool-use plan}
% \ForAll{$call \in tool_{plan}$}
%     \State $Tool_{selected}, Args \gets call.name, call.arguments$
%     \State $result \gets \text{MCP.CallTool}(Tool_{selected}, Args)$ 
%     \State Update execution context with $result$
% \EndFor
% \State $R_{final} \gets \text{LLM.SynthesizeResponse}(Q, \text{execution context})$ \Comment{Synthesize final output}
% \State \Return $R_{final}$
% \end{algorithmic}
% \end{algorithm}

% This architecture's capabilities are illustrated by a physically embodied strategy-board assistant, encapsulated as a single neuro-symbolic tool within the MCP ecosystem. When a user requests move suggestions in a chess-style game, the assistant executes a hybrid pipeline: (1) \textbf{neuro-symbolic perception} maps the live camera view of the board to a stable symbolic state representation (e.g., a FEN string); (2) \textbf{symbolic search} feeds this structured state into a deterministic engine that evaluates candidate moves; and (3) \textbf{semantic interpretation} uses the orchestrator LLM to translate low-level coordinate outputs (such as ``e2e4'') into high-level strategic commentary that is understandable to non-expert players.

This architecture's capabilities are illustrated by a physically
embodied strategy-board assistant, encapsulated as a single
neuro-symbolic tool within the MCP ecosystem. When a user requests
move suggestions, the assistant executes a hybrid pipeline:
(i) a perception module maps the egocentric board view to a stable
symbolic state (e.g., FEN); (ii) a deterministic engine
evaluates candidate moves; and (iii) the orchestrator LLM translates
coordinate outputs into strategic commentary understandable to
non-expert players.
To ensure robustness against detection noise, we implement a temporal buffer mechanism that commits to a board state only after a stability threshold is met via majority voting. The stabilized state is then processed by the symbolic engine, and the result is synthesized by the LLM. Formal definitions of this smoothing mechanism and the complete execution algorithm are provided in Appendix~\ref{apx:extended_methodology}.

This modular, high-level abstraction enables the orchestration language model to leverage the mature symbolic engine without managing its internal mechanics.
Implementation details of the vision model, state-stabilization heuristics, and egocentric prompting are summarized in Appendix~\ref{apx:extended_methodology}. As with other tools in our system, the final textual response can be synthesized into expressive speech, and the entire end-to-end process is managed by an asynchronous event loop to preserve responsiveness during real-time interaction.

\subsection{On-Device Perception and WebRTC-Based Interaction}
\label{subsec:ondevice_webrtc}

% Our system's front-end architecture, designed for resource-constrained smart glasses, handles real-time, bidirectional multimodal communication with a cloud-native backend. On device, concurrent audio and visual pipelines run in a single event loop. The audio side uses a lightweight, amplitude-based VAD with a short pre-roll buffer, so that full utterances are captured while still supporting full-duplex barge-in: users can speak over system playback, and interruptions are forwarded immediately. In parallel, the video pipeline crops and downscales the egocentric stream before encoding frames for transmission. Let $o_t = (a_t, v_t)$ denote the audio and video observations at time $t$; the front end implements a perception--interaction operator $\mathcal{I}_{\text{front}}$ that maps the history $o_{1:t}$ into streamed media and control signals sent to the backend.

Designed for resource-constrained smart glasses, our front-end handles real-time bidirectional multimodal communication with a cloud backend.
On device, concurrent audio and video
pipelines run in a single event loop. A lightweight energy-based VAD
with a short pre-roll buffer captures full utterances while supporting
full-duplex barge-in, and the video pipeline crops and downscales
the egocentric stream before encoding it for transmission.

\begin{figure}[!t]
    \centering
    \includegraphics[width=0.99\linewidth]{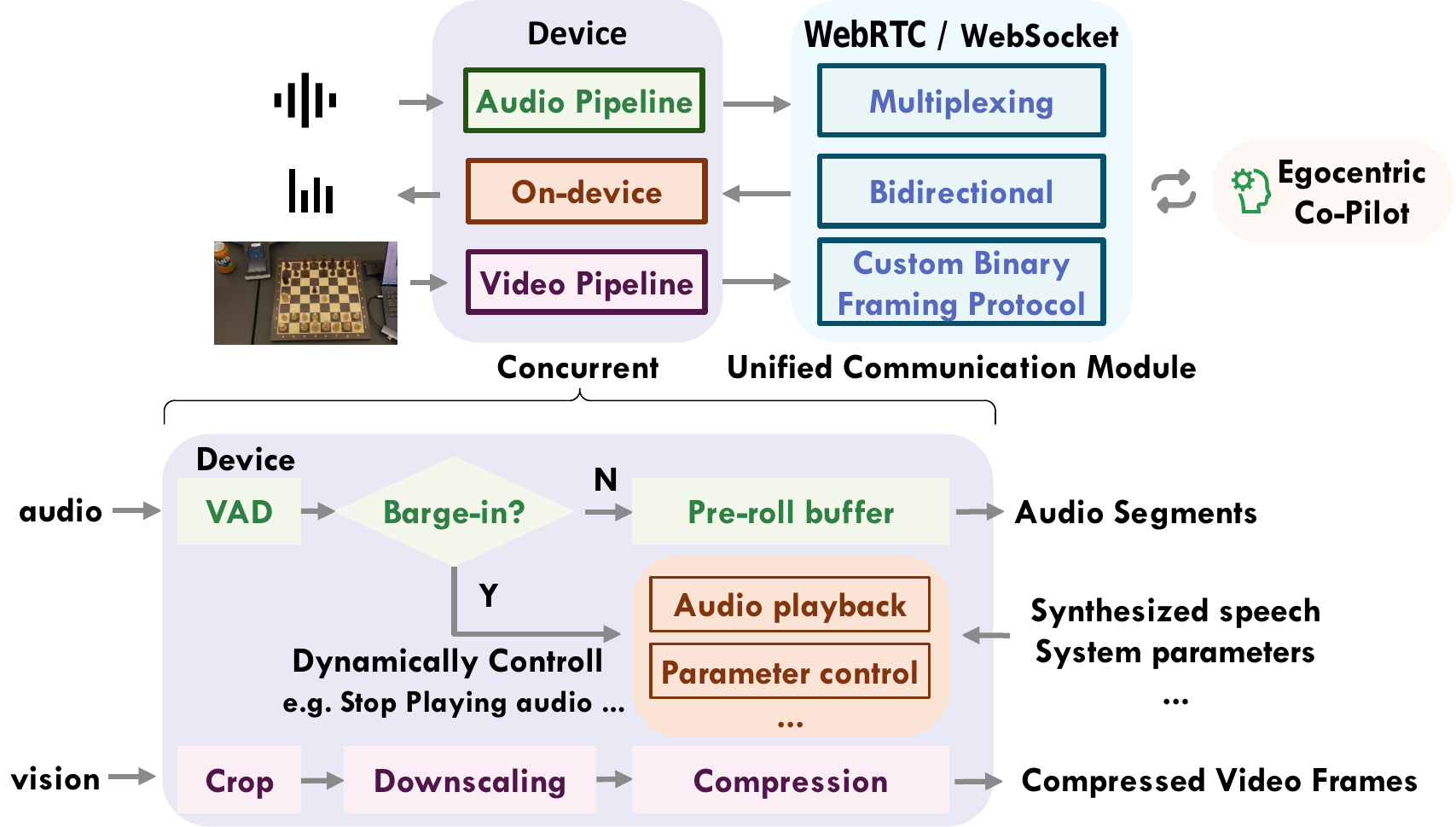}
    % \caption{On-device architecture for real-time multimodal interaction  $\mathcal{I}$. Top: audio and video are processed concurrently and multiplexed into a single bidirectional channel between smart glasses and the cloud-based Egocentric Co-Pilot. In our main deployment this channel is implemented via WebRTC \cite{DBLP:conf/mipro/SredojevSP15} (with H.264 video, Opus audio, and a data channel), while a custom WebSocket protocol serves as an on-premise baseline. Bottom: a detailed view of the audio and video pipelines. The audio path combines energy-based VAD, a pre-roll buffer, and barge-in control; the video path performs cropping, downscaling, and compression, with task-dependent frame rate and quality.}
    % \caption{On-device architecture for real-time multimodal interaction. Audio and video are processed in parallel and multiplexed into a single bidirectional channel between smart glasses and the cloud backend. In our main deployment this channel is implemented via WebRTC \cite{DBLP:conf/mipro/SredojevSP15} (H.264 video, Opus audio, and a data channel), while a custom WebSocket variant serves as an on-premise baseline.}
    \caption{On-device architecture for real-time multimodal interaction. Audio and video are parallel-processed and multiplexed into a bidirectional channel between smart glasses and the cloud backend. We use WebRTC \cite{DBLP:conf/mipro/SredojevSP15} (H.264 video, Opus audio, data channel), with a custom WebSocket variant as an on-premise baseline.}
    \label{fig:ondevice-arch}
\end{figure}

% Instead of a custom WebSocket framing protocol, we employ a standard WebRTC stack (via LiveKit \cite{livekit-github}) to transport both media and control messages in our primary system. Audio is streamed as Opus, video as H.264, and a data channel is used for low-rate JSON control messages and alignment metadata. On the server side, a single voice-pipeline agent composes neural VAD, streaming ASR, a multimodal LLM, and TTS,
% \begin{equation}
%   \mathcal{I}_{\text{webrtc}} 
%   = \mathcal{F}_{\text{vad}} \circ \mathcal{F}_{\text{asr}} \circ \mathcal{F}_{\text{llm}} \circ \mathcal{F}_{\text{tts}},
% \end{equation}
% and is configured for full-duplex operation with interruption-aware endpointing. Smart glasses and a browser-based playground both obtain short-lived access tokens from a lightweight web service and then join the same WebRTC room, reusing the identical operator $\mathcal{I}_{\text{webrtc}}$ to achieve sub-second, web-native multimodal interaction. In our experiments we also deploy an on-premise WebSocket variant, which trades some deployment simplicity and mobility for minimal latency and offers a contrasting point in the design space of energy- and resource-aware web agents.

Instead of a custom WebSocket framing protocol, we employ a standard
WebRTC stack (via LiveKit~\cite{livekit-github}) to transport audio, video, and
low-rate JSON control messages over a unified channel. Audio is
streamed as Opus, video as H.264, and a data channel carries alignment
metadata and tool-calling signals. On the server side, a single
voice-pipeline agent composes neural VAD, streaming ASR, a multimodal
LLM, and TTS, formulated as
\begin{equation}
  \mathcal{I}_{\text{webrtc}} 
  = \mathcal{F}_{\text{vad}} \circ \mathcal{F}_{\text{asr}} \circ \mathcal{F}_{\text{llm}} \circ \mathcal{F}_{\text{tts}},
\end{equation}
and is configured for full-duplex, interruption-aware operation.
Smart glasses and a browser-based playground reuse the same operator
$\mathcal{I}_{\text{webrtc}}$ to achieve sub-second, web-native multimodal
interaction. For comparison we also deploy an on-premise WebSocket
variant, which trades some deployment simplicity and mobility for
slightly lower latency.

\subsection{Proactive Multimodal Intent Disambiguation}
\label{subsec:clarifier}

Even with a strong multimodal backbone, users frequently issue
underspecified or ambiguous instructions in egocentric settings
(e.g., “Can you show me this again?” while pointing at a board or
appliance). To mitigate misunderstandings—which can be especially
harmful in assistive or educational contexts—we integrate a lightweight,
plug-and-play clarifier at the end of the interaction pipeline.
When the LLM detects high semantic uncertainty or conflicting
interpretations, the clarifier reframes the situation as a constrained
decision problem. Given an input $(x_{1:t}, v_{1:t}, s_t)$ and a small
set of candidate interpretations $\{\phi_k\}$, it chooses between
answering directly or asking a short clarification question:
\begin{equation}
\phi^\star = \arg\max_{\phi_k \in \Phi \cup \{\text{ask}\}}
  U\big(\phi_k \mid x_{1:t}, v_{1:t}, s_t\big),
\end{equation}
where $U(\cdot)$ trades off informativeness and interaction cost.
If $\phi^\star = \text{ask}$, the system issues a brief follow-up
(e.g., ``Do you mean the piece on the left or the one near the
corner?'') and updates the context with the reply before committing
to an action. We instantiate this module by adapting a recently
proposed plug-and-play multimodal clarifier~\cite{yang2026plugandplayclarifier}, and focus
here on its integration into an egocentric, WebRTC-based assistant.
The clarifier acts as a black-box wrapper around the underlying
LLM and perception operators, requires no retraining of foundation
models, and can be selectively enabled for sensitive domains such
as navigation aids or daily assistance.

\paragraph{Runtime Guardrails and Schema Management.}

While Algorithm~\ref{alg:generalized_loop} presents a linear plan--then--execute loop for clarity, our implementation includes pragmatic guardrails to prevent unsafe or unintended actions. 
First, tools are exposed to the LLM through an explicit allowlist, and in all smart-glasses experiments we restrict the tool set to non-destructive capabilities (e.g., query answering) without direct control over actuators or external accounts. Second, before each tool call, MCP validates arguments against a type-annotated schema derived from the function signature and a structured docstring; mismatches are logged and the call is aborted rather than coerced. Third, for commands that could have side effects (such as editing a calendar entry), the orchestrator is required to issue a natural-language confirmation prompt, and the tool is only executed when the user explicitly confirms the action. Multi-tool plans are executed in a best-effort manner: if any intermediate call fails, the remaining calls are skipped and the LLM is instructed to summarize the partial result instead of attempting an automatic rollback. In this work we focus on the reasoning and orchestration aspects; industrial deployments would additionally require stronger mechanisms such as capability whitelists per application, schema versioning, and transactional commit/abort semantics.
% perception, query answering, local note-taking
\section{Experiments}

\subsection{Application to Egocentric QA Benchmarks}
\label{sec:implementation_details}

We apply our Reasoning Core to the Egolife and HD-EPIC benchmarks, each of which presents unique challenges. For the long-form videos in Egolife, which require reasoning over extensive history, we use {Hierarchical Context Compression (HCC)}. Specifically, the historical log is divided into temporal chunks (e.g., hourly). A text-only LLM then evaluates the relevance of each chunk to the user's query and generates a concise summary for only the relevant ones. This process produces a compact, query-specific representation of the past, which is prepended to a detailed log of recent events to create the final context for the reasoning MLLM. In contrast, for the action-focused clips in HD-EPIC, we use specialized {Temporal Chain-of-Thought (T-CoT)} strategies. To reason about multiple clips, relevant video segments are programmatically joined into a single timeline with re-normalized timestamps. For single videos that exceed the context window, we generate a textual summary by describing sequential segments, which is then used as context. For all benchmarks, we use a two-stage process for robust answer generation: a regex-based parser first extracts the primary answer choice, followed by a majority vote over the outputs from five syntactically different prompts.

Table \ref{tab:my-table-1} shows our main results on the Egolife and HD-EPIC QA benchmarks and compares the performance of our Reasoning Core against state-of-the-art methods. Our approach achieves strong results, particularly on HD-EPIC, which highlights the utility of our dynamic T-CoT strategies for action-centric reasoning.

\begin{table}[!t]
\centering
\small
\setlength{\tabcolsep}{6pt}
\renewcommand{\arraystretch}{1.1}
\begin{tabular}{llc}
\toprule
\tblhead
\textbf{Dataset} & \textbf{Model} & \textbf{Accuracy (\%)} \\
\midrule
\multirow{5}{*}{\makecell[c]{Egolife\\\cite{yang2025egolife}}}
  & LLaVA-OV~\cite{DBLP:journals/tmlr/0080ZGZ00ZZL0L25}       & 30.8\textsuperscript{*} \\
  & GPT-4o~\cite{DBLP:journals/corr/abs-2410-21276}           & 36.2\textsuperscript{*} \\
  & Gemini-1.5-Pro~\cite{DBLP:journals/corr/abs-2403-05530}   & 36.9\textsuperscript{*} \\
  & Qwen2.5 VL~\cite{DBLP:journals/corr/abs-2502-13923}       & 38.1\textsuperscript{\,} \\
  & \textbf{Ours}                                             & \textbf{40.9}\textsuperscript{\,} \\
\midrule
\multirow{6}{*}{\makecell[c]{HD-EPIC\\\cite{DBLP:conf/cvpr/PerrettDSEPPLGB25}}}
  & VideoLlama 2~\cite{DBLP:journals/corr/abs-2406-07476}     & 27.4\textsuperscript{*} \\
  & LongVA~\cite{DBLP:journals/corr/abs-2406-16852}           & 29.3\textsuperscript{*} \\
  & LLaVA-Video~\cite{DBLP:conf/emnlp/LinYZCNJ024}            & 32.4\textsuperscript{*} \\
  & Qwen2.5 VL~\cite{DBLP:journals/corr/abs-2502-13923}       & 33.5\textsuperscript{\,} \\
  & Gemini-1.5-Pro~\cite{DBLP:journals/corr/abs-2403-05530}   & 37.6\textsuperscript{*} \\
  & \textbf{Ours}                                             & \textbf{46.2}\textsuperscript{\,} \\
\bottomrule
\end{tabular}
\caption{Comparison against state-of-the-art methods on the Egolife and HD-EPIC benchmarks. Results marked with an asterisk (*) are reported in the original papers; all other results are from our reproductions using official code.}
\label{tab:my-table-1}
\end{table}

\subsection{Ablation and Sensitivity Analysis}

To clarify the contribution of each component, we conduct ablations summarized in Table~\ref{tab:ablation_results} and highlight the key findings here. On {Egolife}, removing HCC reduces accuracy by 2.0 points, while removing T-CoT yields a 1.4-point drop, confirming that both long-horizon summarization and local temporal structuring contribute meaningfully. Fine-tuning on egocentric data accounts for a further 1.7-point gain, and dropping ASR transcripts costs 0.8 points, indicating that spoken cues provide useful but secondary context. On {HD-EPIC}, which emphasizes short but complex action clips, domain-specific fine-tuning is even more critical: omitting it leads to a 5.62-point degradation. Removing HCC and T-CoT reduces accuracy by 4.68 and 3.55 points respectively, showing that temporal organization still matters even at clip scale. Finally, disabling our prompt- and output-hygiene layer (pre-processing, regex answer extraction, and prompt ensembling) yields a 2.67-point drop, so seemingly ``engineering'' details are empirically important for robustness.
% \begin{table}[!t]
% \centering
% \begin{tabular}{clc}
% \toprule
% \textbf{Dataset} & \multicolumn{1}{c}{\textbf{Variant}} & \textbf{Accuracy (\%)} \\
% \midrule
% \multirow{5}{*}{\makecell[c]{Egolife\\\cite{yang2025egolife}}}
% & Ours & \textbf{40.9} \\
% & w/o Fine-tuning & 39.2 (-1.7) \\
% & w/o T-CoT & 39.5 (-1.4) \\
% & w/o HCC & 38.9 (-2.0) \\
% & w/o Transcript & 40.1 (-0.8) \\
% \midrule
% \multirow{5}{*}{\makecell[c]{HD-EPIC\\\cite{DBLP:conf/cvpr/PerrettDSEPPLGB25}}} 
% & Ours & \textbf{46.23} \\
% & w/o Fine-tuning & 40.61 (-5.62) \\
% & w/o T-CoT & 42.68 (-3.55) \\
% & w/o HCC & 41.55 (-4.68) \\
% & w/o Pre-Processing & 43.56 (-2.67) \\
% \bottomrule
% \end{tabular}
% \caption{Ablation study of our Reasoning Core on Egolife and HD-EPIC. We report the full model (``Ours'') and variants where a single component is removed; the performance drop is shown in parentheses.}
% \label{tab:ablation_results}
% \end{table}
\begin{table}[!t]
\centering
\small
\setlength{\tabcolsep}{6pt}
\renewcommand{\arraystretch}{1.1}
\begin{tabular}{llc}
\toprule
\rowcolor{black!5}
\textbf{Dataset} & \textbf{Variant} & \textbf{Accuracy (\%)} \\
\midrule
\multirow{5}{*}{\makecell[c]{Egolife\\\cite{yang2025egolife}}}
  & Ours              & \textbf{40.9}          \\
  & w/o Fine-tuning   & 39.2 (-1.7)            \\
  & w/o T-CoT         & 39.5 (-1.4)            \\
  & w/o HCC           & 38.9 (-2.0)            \\
  & w/o Transcript    & 40.1 (-0.8)            \\
\midrule
\multirow{5}{*}{\makecell[c]{HD-EPIC\\\cite{DBLP:conf/cvpr/PerrettDSEPPLGB25}}}
  & Ours              & \textbf{46.23}         \\
  & w/o Fine-tuning   & 40.61 (-5.62)          \\
  & w/o T-CoT         & 42.68 (-3.55)          \\
  & w/o HCC           & 41.55 (-4.68)          \\
  & w/o Pre-Processing& 43.56 (-2.67)          \\
\bottomrule
\end{tabular}
\caption{Ablation study of our Reasoning Core on Egolife and HD-EPIC. We report the full model (``Ours'') and variants where a single component is removed; the performance drop is shown in parentheses.}
\label{tab:ablation_results}
\end{table}
We also analyze HCC sensitivity by varying chunk size and summary length: halving summary length or doubling chunk size reduces Egolife accuracy by at most 1.0--1.6 points, while switching the selection LLM to a smaller model causes a 2.6-point drop. 
% These trends indicate that the framework is robust to reasonable hyperparameter changes, and that HCC provides benefits beyond what T-CoT alone can offer.
These trends suggest that the framework is reasonably robust to hyperparameter changes, and that HCC adds value on top of T-CoT alone.
\begin{figure*}[!t]
    \centering
    \includegraphics[width=0.78\textwidth]      {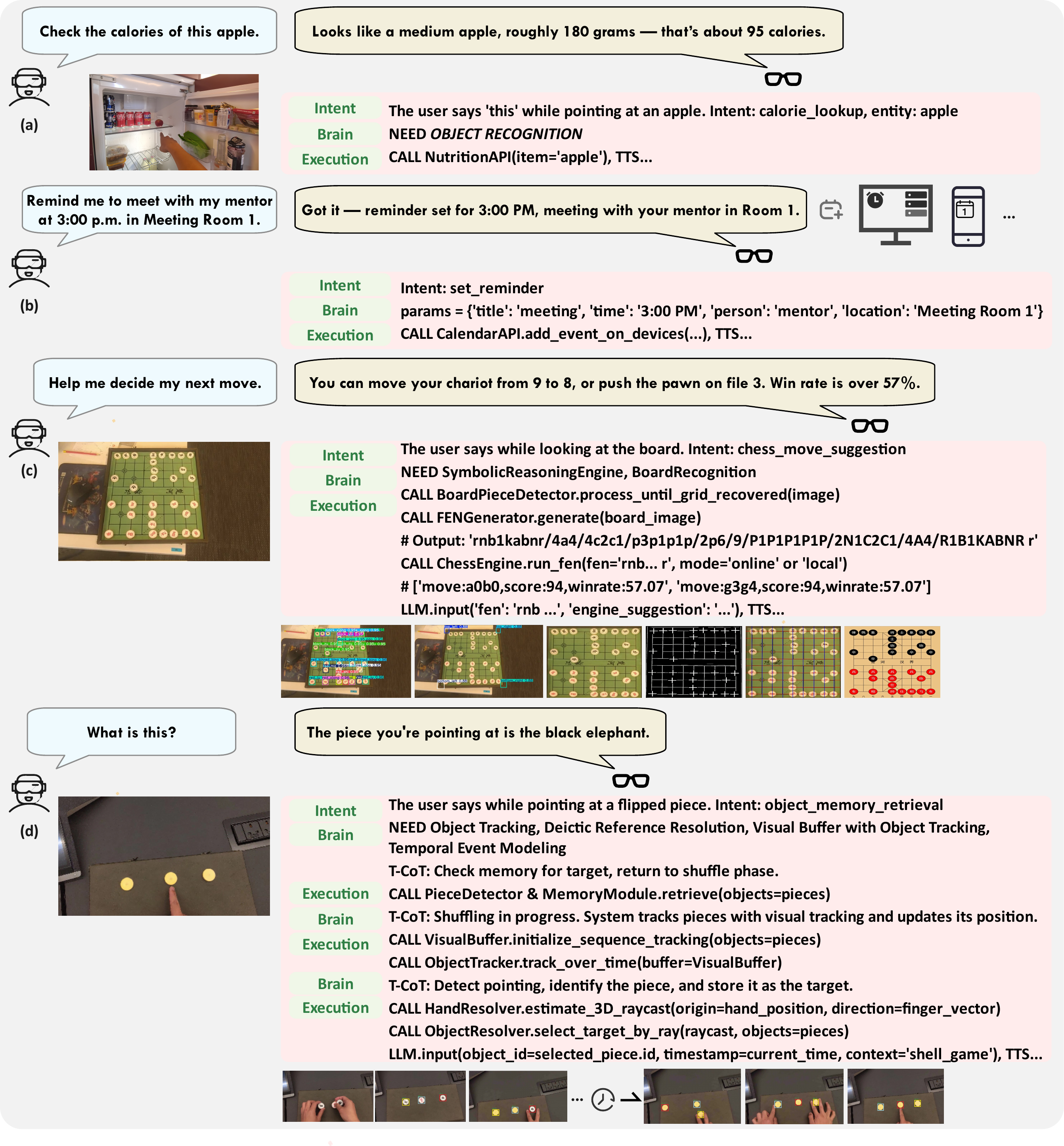}      % 775
    \caption{Core capabilities of the LLM-orchestrated framework. Our system interprets multimodal user intent and dynamically composes neuro-symbolic tools via the MCP protocol. (a) Foundational Tool Use: a simple query triggers a VLM for object recognition and an external API call. (b) Structured Task Management: natural language is translated into a structured API call for a native device application. (c) Complex Neuro-Symbolic Reasoning: the board-game co-pilot integrates a vision tool (neuro), a deterministic game engine (symbolic), and an LLM for semantic explanation. (d) Spatiotemporal Memory: the system resolves a deictic reference (``this'') by visually tracking an object through occlusion and recalling it from memory.}
    \label{fig:tool_pipeline}
\end{figure*}

\subsection{Tool Use in Real-World Scenarios}
\label{sec:tools_experiments}

To validate our LLM-orchestrated neuro-symbolic framework, the Egocentric Co-Pilot, we conducted real-world egocentric experiments on an end-to-end system running on smart glasses. Prompted by user commands in English or Mandarin, these experiments evaluated the system's ability to interpret multimodal intent, compose tools, and execute complex tasks. We measured performance using the Task Completion Rate (TCR), defined as successful task execution without user intervention, with further details on configuration summarized in Appendix~\ref{apx:extended_methodology}.
We organized tasks into three categories to reflect increasing system complexity and real-world risk. Category~1 (Foundational Tool Use) contains frequent, low-risk requests such as fact lookup, simple reminders, and note-taking; these probe whether natural language can be reliably mapped to web APIs and local utilities. Category~2 (Embodied and Spatiotemporal Tasks) focuses on perceptually grounded activities, such as over-the-board game assistance and object tracking, which require stable egocentric perception and short-term memory. Category~3 (Complex Neuro-Symbolic Reasoning) groups expert-style tasks that combine noisy visual input with deterministic symbolic solvers. In total we define several dozen unique task templates, with a roughly balanced distribution across the three categories; each template is instantiated into multiple concrete trials during evaluation. This hierarchy allows us to separately stress-test core API grounding, embodied perception, and full neuro-symbolic reasoning, while covering both common daily needs and more demanding long-tail scenarios such as strategy tutoring.

\paragraph{Category 1: Foundational Tool Use.}
This category tests the core ability to map natural language to specific API calls. Tasks include querying knowledge bases (e.g., ``Check the calories of this apple''), managing personal information (e.g., ``Remind me of the 3 PM meeting''), or creating notes. As illustrated in Figure~\ref{fig:tool_pipeline}(a--b), these tasks require the LLM to parse intent, extract entities (sometimes from visual context), and invoke the correct tool (e.g., \texttt{NutritionAPI}, \texttt{CalendarAPI}, \texttt{MemoTool}) with proper arguments. The high TCR of {98.5\%} across these tasks demonstrates the reliability of our fundamental execution loop.

\paragraph{Category 2: Embodied Strategy and Spatiotemporal Tasks.}
This category is instantiated by an over-the-board strategy assistant that operates on chess-style games played on a physical board. The assistant is wrapped as a single neuro-symbolic tool in the MCP ecosystem. When the user asks for move suggestions, the tool runs the hybrid pipeline summarized in Algorithm~\ref{alg:chess_tool}: a vision module observes the current board position and converts raw frames into a stable symbolic state (Eq.~\ref{eq:chess_smoothing}); a deterministic engine then performs symbolic search over legal moves; finally, the orchestration LLM turns coordinate-style outputs into strategic natural-language guidance tailored to the player's skill level. This design exemplifies embodied, spatiotemporal reasoning: the task is grounded in a continuously evolving physical scene, yet the assistant communicates through speech and text. By exposing only a clean tool interface to the MCP orchestrator, we allow the language model to call into a sophisticated game engine without being entangled with its internal logic, while still providing real-time, multimodal feedback to the user via synthesized speech.

\paragraph{Category 3: Complex Neuro-Symbolic Reasoning.}
% This final category serves as the capstone evaluation, assessing the entire neuro-symbolic pipeline on tasks that demand a tight coupling of real-world perception with formal symbolic reasoning. 
This final category evaluates the entire neuro-symbolic pipeline on tasks that demand a tight coupling of real-world perception with formal symbolic reasoning.
Such tasks are characterized by the need to (i) convert noisy visual input into a structured, symbolic representation, (ii) apply a deterministic or heuristic rule-based engine to this representation, and (iii) translate the symbolic output back into contextually aware, natural-language guidance. We use a board-game co-pilot instantiated on several chess-style games as a representative benchmark for this category. 
% Across 50 games, the system achieved an end-to-end success rate of 98\% in generating strategically sound and contextually relevant move suggestions, validating the seamless integration of its robust perception module, the symbolic engine, and the LLM-driven semantic interpreter as formalized in Algorithm~\ref{alg:chess_tool}.
Across 50 games, the system achieved an end-to-end success rate of 98\% in generating strategically sound and contextually relevant move suggestions, illustrating that the perception module, the symbolic engine, and the LLM-driven semantic interpreter work well together as formalized in Algorithm~\ref{alg:chess_tool}.
\paragraph{Failure Analysis.}

To better understand the limitations of Egocentric Co-Pilot, we manually inspected a representative set of failure cases across all three task categories. Most failures fell into four buckets: (i) perception errors, such as mis-detected board states or mislocalized target objects in cluttered scenes; (ii) intent misunderstandings, where the LLM overgeneralized from context and chose an incorrect tool or misinterpreted a deictic reference; (iii) tool-level issues, including missing arguments or unexpected API responses; and (iv) long-horizon memory errors, where relevant past events were omitted from the compressed context. Perception and intent errors were the most common, especially under poor lighting or rapid head motion. 
These categories suggest concrete mitigations such as stronger egocentric backbones, explicit confirmation turns in ambiguous situations, and stricter argument validation. We leave a systematic exploration of these directions to future work.
Across categories, we deliberately focus on tasks that reflect constructive everyday assistance---such as reading labels, managing simple schedules, and receiving over-the-board tutoring---rather than entertainment-oriented or engagement-only scenarios. This choice is intended to better capture how such agents can support users in practical daily activities that affect autonomy and well-being.

\subsection{Human-in-the-Loop Evaluation}
\label{sec:human_eval}

To assess real-world efficacy, we conducted a human-in-the-loop study comparing our Egocentric Co-Pilot against several commercial smart-glasses devices and a human baseline. Rather than running live interactive sessions---which would confound AI quality with hardware, network, and connectivity differences---we adopted a controlled offline protocol: for each system and each task, we recorded interaction logs (audio, video, and transcripts) using identical prompts and environments, then asked independent raters to evaluate these logs.

\begin{figure}[!t]
\centering
\includegraphics[width=0.995\columnwidth]{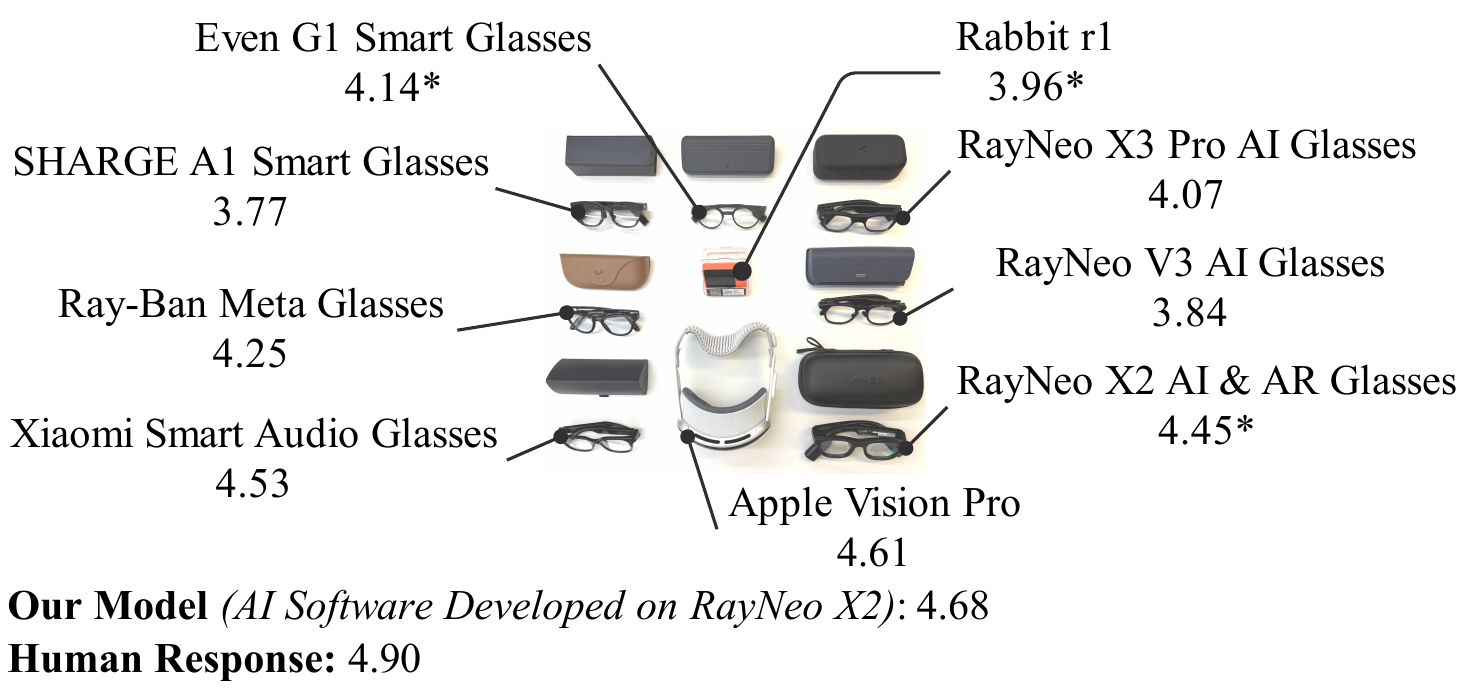}
% \caption{Subjective evaluation of our Egocentric Co-Pilot against commercial devices and a human baseline. Scores represent mean user ratings on a 5-point Likert scale, where higher is better. Our model significantly outperforms all commercial competitors, validating its superior real-world performance. Asterisks (*) denote devices with non-standard AI interaction patterns.}
\caption{Subjective evaluation of Egocentric Co-Pilot against commercial smart-glasses devices and a human baseline. Bars show mean 5-point Likert ratings (higher is better); asterisks (*) denote devices whose default interaction pattern deviates from continuous conversational AI.}
\label{fig:human_eval}
\end{figure}

Participants were presented with anonymized, randomly ordered clips and were blinded to the identity of the underlying system. For each clip, they rated on a 5-point Likert scale (higher is better) whether the assistant (i) correctly understood multimodal intent and (ii) successfully executed the corresponding task. We additionally recorded an objective Task Completion Rate (TCR), defined as a task completed without human intervention according to a pre-defined success checklist. Devices whose default interaction pattern deviates substantially from continuous conversational AI (e.g., notification-only modes) are marked with an asterisk in Figure~\ref{fig:human_eval}; we still include them as baselines but interpret their scores with this caveat in mind.

As shown in Figure~\ref{fig:human_eval}, our model, deployed on standard off-the-shelf hardware, achieved a mean rating of 4.70, significantly surpassing all commercial competitors and approaching the human baseline of 4.92. Average TCR followed a similar trend, with Egocentric Co-Pilot completing more tasks end-to-end than any individual device baseline. These gains align with our design goals: improved intent disambiguation and robust tool composition translate into fewer user corrections and more satisfying assistance. Because this study involved rating pre-recorded, fully anonymized interaction logs for low-risk daily tasks (e.g., weather queries, simple planning, over-the-board game advice), it fell under the ``minimal risk'' category at our institution and did not require formal IRB review; all participants gave informed consent prior to participation. Additional protocol details are summarized in Appendix~\ref{sec:appendix_human_eval}, and please see Figure \ref{fig:human_logs} for an example.
These results suggest that, even on commodity hardware, a carefully orchestrated web-native assistant can provide users with more reliable, less frustrating support for everyday tasks than current commercial smart-glasses software, pointing toward a practical path for deploying egocentric web agents that genuinely improve day-to-day autonomy rather than merely adding notifications.

\section{Limitations and Future Work}
Egocentric Co-Pilot is a research prototype with several limitations. First, its behavior ultimately depends on the underlying LLM/VLM backbones and on hand-designed tool schemas. Errors in perception, reasoning, or tool selection can still cascade through the pipeline, and our current guardrails (allowlisted tools, schema-based argument checks, and explicit confirmations) are weaker than formal safety guarantees. Extending the framework with stronger capability management, per-application policies, and transactional commit/abort semantics is an important direction.

Second, our reliance on domain-adapted egocentric models and a cloud backend. While fine-tuning on first-person data improves performance, it may not transfer perfectly to new domains or camera form factors, and continuous streaming introduces latency and energy costs. We plan to explore parameter-efficient adaptation, capable on-device models to reduce streaming, and explicit accounting of compute and energy footprints across deployment options.

Finally, our evaluation focuses on short-term assistance and strategy tutoring with healthy adults in controlled settings. We exclude long-term effects, high-stakes scenarios, or the needs of people with disabilities, older adults, or other groups who might benefit most. Privacy and bystander consent also remain open concerns for always-on egocentric capture. Future work includes longitudinal studies with diverse populations and stronger on-device filtering and privacy-preserving training tailored to web-scale deployment.

\section{Conclusion}
We introduce Egocentric Co-Pilot, a modular neuro-symbolic framework integrating egocentric perception, long-horizon context management, and LLM-orchestrated tool use in a single smart-glasses assistant. Combining Temporal Chain-of-Thought and Hierarchical Context Compression with a web-native tool ecosystem and a cloud-native WebRTC backend, it delivers competitive accuracy on Egolife and HD-EPIC and
outperforms several commercial assistants in real-world human-in-the-loop studies.
% , in a human-in-the-loop study, outperforms several commercial smart-glasses assistants on real-world tasks. 

Beyond raw performance, the design emphasizes assistive, everyday use cases such as situated tutoring, context-aware reminders, and reading support, aiming to enhance independence and digital well-being rather than optimize engagement alone. We hope that Egocentric Co-Pilot can serve as a concrete blueprint for future web-native egocentric agents that are not only technically capable, but also deployable as responsible, inclusive technologies for people who stand to benefit most from contextual, always-on assistance.
More broadly, our results suggest that carefully orchestrating specialized tools around a principled sensing and context-management stack can be a more practical path toward trustworthy, assistive AI on the web than simply scaling monolithic models.

\section{Acknowledgments}

This work was supported by the Shenzhen Science and Technology Program (Grant No. ZDSYS20220323112000001).

% \clearpage

\bibliographystyle{ACM-Reference-Format}
\bibliography{sample-base}

% \clearpage

\appendix

\section{Extended Methodology}
\label{apx:extended_methodology}

For completeness, we briefly summarize several implementation details of the Egocentric Co-Pilot that complement the main text. The unified event log $\mathcal{E}$ is constructed by sampling egocentric video at 1~FPS and running a fine-tuned \texttt{Qwen2.5-VL-7B-Instruct} model \cite{DBLP:journals/corr/abs-2502-13923} to generate dense, first-person descriptions of actions, object state changes, and scene context. These visual entries are merged with ASR transcripts into a single, time-ordered sequence of events. Queries are pre-processed by analyzing their modality requirements (image, video, or mixed), rewriting under-specified questions into explicit, viewpoint-grounded prompts, and reformatting multiple-choice options into a consistent template; this reduces parsing ambiguity and improves robustness.

The egocentric backbone is obtained by fine-tuning {Qwen2.5-VL-7B-Instruct}  \cite{DBLP:journals/corr/abs-2502-13923} on a mixture of first-person video datasets, % \texttt
including EPIC-KITCHENS \cite{DBLP:journals/ijcv/DamenDFFKMMMPPW22, DBLP:journals/corr/abs-1804-02748}, EgoProceL \cite{DBLP:conf/eccv/BansalAJ22}, YOUCOOK2 \cite{DBLP:conf/aaai/ZhouXC18}, VISOR \cite{DBLP:conf/nips/DarkhalilSZMKHF22}, EgoIT \cite{yang2025egolife}, and relevant portions of Ego4D \cite{DBLP:conf/cvpr/GraumanWBCFGH0L22}. We freeze the vision tower and projector, update LLM layers with AdamW (learning rate $2 \times 10^{-7}$, batch size 2, one epoch, bfloat16 precision), and cap both frame count and sequence length (up to 131{,}072 tokens). Temporal Chain-of-Thought (T-CoT) is implemented via simple prompt templates that encourage intermediate reasoning steps and by programmatically cropping or concatenating temporal windows, as described in Section~\ref{sec:brain}. Post-processing consists of regular-expression extraction of answer letters and majority voting over five syntactically distinct prompts per question.

\subsection*{Generalized Execution Loop}
\label{apx:gen_loop}
The generalized orchestration loop used by the Egocentric Co-Pilot is detailed in Algorithm~\ref{alg:generalized_loop}.

\begin{algorithm}[t]
\small
\caption{Generalized LLM-Orchestrated Execution Loop}
\label{alg:generalized_loop}
\begin{algorithmic}[1]
\Require User Query $Q$, Multimodal Context $C_{mm}$
\State $T_{available} \gets \text{MCP.ListTools}()$ \Comment{Discover available tools}
\State $tool_{plan} \gets \text{LLM.GeneratePlan}(Q, C_{mm}, T_{available})$ \Comment{LLM formulates a tool-use plan}
\ForAll{$call \in tool_{plan}$}
    \State $Tool_{selected}, Args \gets call.name, call.arguments$
    \State $result \gets \text{MCP.CallTool}(Tool_{selected}, Args)$ 
    \State Update execution context with $result$
\EndFor
\State $R_{final} \gets \text{LLM.SynthesizeResponse}(Q, \text{execution context})$ \Comment{Synthesize final output}
\State \Return $R_{final}$
\end{algorithmic}
\end{algorithm}

\subsection*{Board-Game Tool Implementation Details}
The board-game co-pilot serves as a prime example of our neuro-symbolic tool usage. To make the perceived board state robust to frame-by-frame detection noise, we maintain a temporal buffer of predictions $V_{r,c,k}^{(i)}$ at each board location $(r,c)$, where $k$ indexes piece types and $i$ indexes frames. The committed state $P_{r,c}$ is obtained by majority vote with a stability threshold $\tau$:
\begin{equation}
    P_{r,c} = 
    \begin{cases} 
        \displaystyle\arg\max_k \sum_{i=1}^N V_{r,c,k}^{(i)} & \text{if } \frac{1}{N}\max_k \sum_{i=1}^N V_{r,c,k}^{(i)} \ge \tau, \\[0.5em]
        P_{r,c}^{\text{prev}} & \text{otherwise,}
    \end{cases}
    \label{eq:chess_smoothing}
\end{equation}
where $P_{r,c}^{\text{prev}}$ is the previously committed state at $(r,c)$, $N$ is the buffer size, and $\tau$ controls the trade-off between responsiveness and stability.

The execution logic is encapsulated in Algorithm~\ref{alg:chess_tool}, demonstrating how the visual stream is converted into natural language advice.

\begin{algorithm}[h] % 注意这里建议改为 [h] 或者 [t]
\small
\caption{Hybrid Neuro-Symbolic Chess-Style Tool Execution}
\label{alg:chess_tool}
\begin{algorithmic}[1]
\Require Visual stream $\mathcal{V}$, LLM $\mathcal{M}_{\mathrm{LLM}}$, symbolic engine $\mathcal{S}_{\mathrm{eng}}$
\Ensure Natural-language strategic advice $\mathcal{A}$
\Function{ExecuteBoardTool}{$\mathcal{V}$}
    \State \textbf{Perception:} $S_{\mathrm{FEN}} \gets \textsc{PerceiveStableState}(\mathcal{V})$ \Comment{Uses Eq.~(\ref{eq:chess_smoothing})}
    \State \textbf{Symbolic search:} $M_{\mathrm{sym}} \gets \mathcal{S}_{\mathrm{eng}}.\textsc{GetBestMove}(S_{\mathrm{FEN}})$
    \State \textbf{Semantic explanation:}
    \State $P \gets$ ``As a board-game coach, explain the idea behind move \texttt{$M_{\mathrm{sym}}$} given the current position.''
    \State $\mathcal{A} \gets \mathcal{M}_{\mathrm{LLM}}.\textsc{Generate}(P)$
    \State \Return $\mathcal{A}$
\EndFunction
\end{algorithmic}
\end{algorithm}

On the tool side, each capability is registered with MCP using a decorator that exposes its type-annotated signature and docstring. The board-game co-pilot described in Algorithm~\ref{alg:chess_tool} uses a compact convolutional network for per-square classification, the temporal smoothing rule of Eq.~\ref{eq:chess_smoothing} to stabilize the perceived position, and a standard chess engine as the symbolic core. In our smart-glasses prototype, the same MCP registry also exposes web APIs (e.g., for weather or nutrition), local utilities (e.g., notes and reminders), and device-bridging tools that send structured JSON messages to nearby phones or computers.

% \section{Algorithmic View of Temporal CoT and HCC}
\section{Real-Time Audio Processing on the Client Device}     % Real-Time 

% Algorithm~\ref{alg:vad} summarizes the query-time
% context construction pipeline combining Temporal Chain-of-Thought
% (T-CoT) and Hierarchical Context Compression (HCC) described in
% Section~\ref{sec:brain}. It makes explicit how the system selects
% relevant temporal windows, summarizes long-horizon history, and
% assembles the final multimodal context for the MLLM.

Algorithm~\ref{alg:vad} details the on-device audio pipeline used in our smart-glasses
prototype. It implements lightweight VAD, pre-roll buffering, and barge-in
detection so that speech segments can be streamed to the cloud with low
latency while still allowing users to interrupt ongoing playback when needed.
This client-side pipeline is used in all our WebRTC-based deployments described in Section~\ref{subsec:ondevice_webrtc}.

\begin{algorithm}[!t]
\small
\caption{Real-time Audio Processing on Client Device}
\label{alg:vad}
\begin{algorithmic}[1]
\State Initialize ring buffer $\mathcal{B}_{\text{ring}}$, state $S \gets \textsc{IDLE}$
\State Define thresholds $\theta_{\text{start}}$, $\theta_{\text{barge-in}}$ and durations $T_{\text{silence}}$, $T_{\min}$
\While{true}
    \State Acquire audio chunk $b$, set $b' \gets g \cdot b$ with $g = 5.0$
    \State Update $\mathcal{B}_{\text{ring}}$ with $b'$, let $A \gets \max(|b'|)$
    \If{system is playing audio \textbf{and} $A > \theta_{\text{barge-in}}$}
        \State \textsc{HaltPlayback}() \Comment{Barge-in detected}
    \EndIf
    \If{$S = \textsc{IDLE}$}
        \If{$A > \theta_{\text{start}}$}
            \State $S \gets \textsc{RECORDING}$
            \State Start new segment with $\mathcal{B}_{\text{ring}}$
        \EndIf
    \ElsIf{$S = \textsc{RECORDING}$}
        \State Append $b'$ to current segment
        \If{$A < \theta_{\text{start}}$}
             \State Start or continue silence timer of length $T_{\text{silence}}$
             \If{timer expired}
                \State Finalize segment $\mathcal{S}_{\text{audio}}$
                \If{$\text{duration}(\mathcal{S}_{\text{audio}}) > T_{\min}$}
                    \State \textsc{Dispatch}$(\mathcal{S}_{\text{audio}})$
                \EndIf
                \State $S \gets \textsc{IDLE}$
             \EndIf
        \Else
            \State Reset silence timer
        \EndIf
    \EndIf
\EndWhile
\end{algorithmic}
\end{algorithm}

\section{Additional Results on Egolife and EgoGPT}
\label{apx:egogpt}

For contextual completeness, we summarize offline EgoGPT results reported in~\cite{yang2025egolife} on the Egolife benchmark. EgoGPT adopts a multi-stage retrieval-augmented generation (RAG+) pipeline with heavy offline processing: it first builds an index over long egocentric video logs and then runs multiple passes of LLM inference over the entire dataset. Under this regime, EgoGPT (EgoIT--Egolife) achieves an accuracy of 38.5\% and EgoGPT (EgoIT) achieves 42.6\% on Egolife~\cite{yang2025egolife}. However, reproducing such a setup in our setting would require running offline inference over the full dataset multiple times, which is incompatible with our focus on low-latency, always-on assistance on smart glasses and with energy-aware deployment considerations. We therefore report these EgoGPT numbers only as contextual references rather than directly comparable baselines, and our main comparisons in Table~\ref{tab:my-table-1} concentrate on single-pass, streaming-friendly systems.

\section{Additional Human Evaluation Details}
\label{sec:appendix_human_eval}

\begin{figure*}[!t]
\centering
\includegraphics[width=0.995\textwidth]{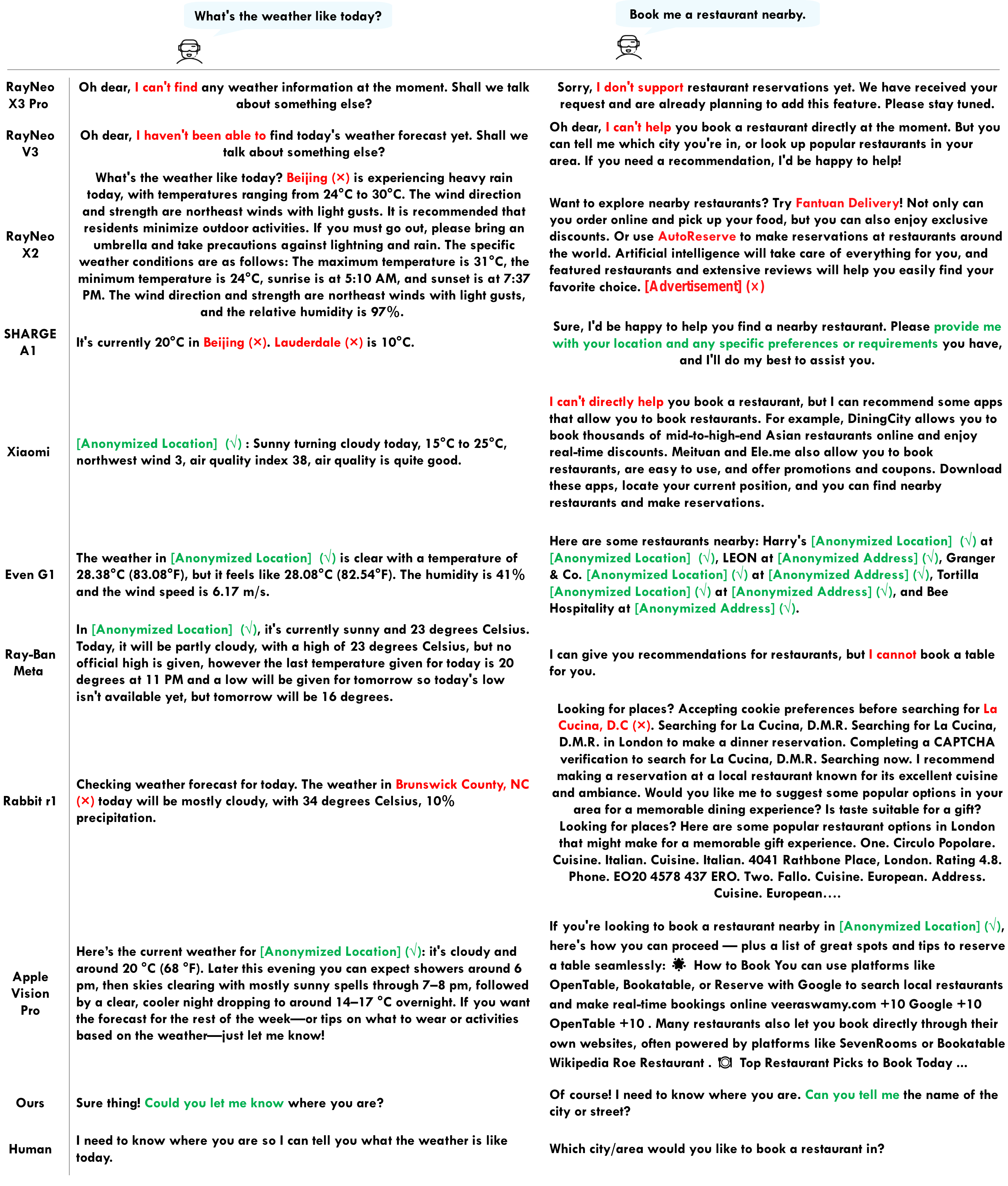}
\caption{Example interaction logs shown to participants. Each column corresponds to a different system's response to the same user query. By evaluating pre-recorded logs instead of live interactions, we avoid confounding AI quality with hardware, network, or UI differences.}
\label{fig:human_logs}
\end{figure*}

The human-in-the-loop study in Section~\ref{sec:human_eval} involved four participants with prior experience using AI assistants. For each task scenario, we collected logs from nine systems: our Egocentric Co-Pilot, several commercial devices, and a human assistant baseline. Logs consisted of audio transcripts, key video frames, and brief textual summaries of system actions (see Figure~\ref{fig:human_logs} for
an example of the anonymized clips shown to participants). All logs were anonymized and shuffled so raters were blind to system identity.
Participants rated each log on two questions using a 5-point Likert scale: (1) how well the assistant understood the multimodal intent, and (2) how successfully it executed the task. The reported scores in Figure~\ref{fig:human_eval} are averages of these two dimensions. Devices with non-standard interaction patterns (e.g., notification-oriented or single-function applications) are marked with an asterisk. Because the study relied on pre-recorded, fully anonymized logs for low-risk daily tasks and did not involve sensitive populations, it was classified as minimal risk under local guidelines; participants provided informed consent and could withdraw at any time without penalty.

\section{Supplementary Discussion on System Deployment and Limitations}
\label{sec:appendix_discussion}

In this section, we provide further details regarding the system's deployment constraints, security considerations, and algorithmic adaptability, addressing specific concerns raised regarding the practical application of egocentric streaming agents.

\subsection{Connectivity and Offline Fallback Strategies}
About the system's behavior in poor network conditions (Offline Fallback). Our current architecture prioritizes a cloud approach due to the strict hardware constraints of wearable AR devices. 
\begin{itemize}
    \item {Hardware Constraints:} The deployment device (RayNeo X2 Pro) imposes significant limitations on Size, Weight, and Power. While we experimented with deploying lightweight models (e.g., 0.5B parameters) locally on the glasses, the inference latency was prohibitive for real-time interaction, and the model capacity was insufficient for complex reasoning.
    \item {Design Choice:} Consequently, we do not currently implement a full offline fallback for complex queries. The system is designed for high-bandwidth environments (WiFi/4G), utilizing the cloud for heavy computation to maintain the wearable form factor. Future iterations may explore hybrid offloading, but currently, stable connectivity is a prerequisite.
\end{itemize}

\subsection{Data Privacy and Security}
Regarding data protection, we acknowledge that this work primarily focuses on the architectural feasibility of egocentric agents.
Standard web protocols are used for transmission. End-to-end encryption and strict data retention policies (e.g., immediate deletion after inference) are planned for the production phase but are not implemented in this prototype.
We propose a "Privacy-First Hybrid Architecture" for future work, where sensitive visual data (e.g., faces, text) is processed or masked locally on the edge device, and only non-sensitive abstract features are transmitted to the cloud.

\subsection{Algorithmic Adaptability: HCC and T-CoT}
We wish to clarify the distinctiveness of our History Context Control (HCC) and Temporal Chain of Thought (T-CoT) compared to standard RAG or Prompt Engineering.
\begin{itemize}
    \item {Online vs. Offline RAG:} Traditional RAG requires offline database indexing, which introduces latency and is ill-suited for the continuous, streaming nature of egocentric video. Our HCC mechanism performs coarse-to-fine retrieval dynamically in the stream, significantly reducing the time compared to retrieving from a static vector database.
    \item {Handling Long Contexts:} Standard MLLMs struggle with the "Lost in the Middle" phenomenon when fed long video histories. Our approach mimics human memory patterns (recency bias) via dynamic compression. The combination of HCC and T-CoT is specifically optimized for the temporal dependencies of first-person video, where understanding the immediate past is often more critical than distant history.
\end{itemize}

\subsection{Scalability of the Toolbox Approach}
The "Toolbox" mechanism is designed as a scalable, hybrid agent system rather than a rigid set of rules.
The system follows a standard agentic paradigm: specific tools (APIs) are defined for high-precision tasks (e.g., Calendar, Weather). However, when a user's intent does not match a predefined tool (or falls into the "long tail" of daily life), the system degrades to the underlying MLLM's general capabilities (Zero-shot VQA). 
% This ensures the system remains functional in chaotic, real-world scenarios without requiring manual engineering for every possible interaction.

\subsection{Hardware Performance: Battery and Thermal Constraints}
% We conducted practical stress tests on the RayNeo X2 Pro to assess battery and thermal performance during continuous streaming.
% \begin{itemize}
The simultaneous operation of the camera, display, and high-frequency network transmission is extremely power-intensive. In continuous streaming mode without external power, the device battery sustains operation for approximately 20 minutes.
We observed that as battery levels drop, the device's firmware triggers power-saving modes that significantly throttle performance (e.g. CPU), causing system lag. 
Thermal dissipation remains within acceptable limits for user comfort, though the device creates noticeable heat during prolonged sessions. These findings reinforce the necessity of our cloud-offloading architecture to minimize on-device compute load, although battery technology remains a bottleneck for all AR hardware.
% \end{itemize}

\end{document}